\documentclass[10pt,twocolumn]{article}

\usepackage[T1]{fontenc}
\usepackage{lmodern}

\usepackage{amsmath,amssymb,amsthm}

\usepackage{graphicx}
\usepackage{float}
\usepackage{booktabs}
\usepackage{multirow}
\usepackage{tabularx}
\usepackage{array}
\newcolumntype{Y}{>{\raggedright\arraybackslash}Y}
\newcolumntype{Z}{>{\centering\arraybackslash}Z}
\renewcommand{\arraystretch}{1.3}
\usepackage{subcaption}

\usepackage{pifont}

\usepackage{enumitem}
\setlist{topsep=3pt, itemsep=3pt, parsep=0pt, partopsep=0pt}
\setlist[enumerate,1]{label=\alph*), leftmargin=*}

\usepackage[numbers,sort&compress]{natbib}

\usepackage[colorlinks=true,linkcolor=blue,citecolor=blue,urlcolor=blue]{hyperref}

\usepackage{geometry}
\usepackage{titlesec}
\titleformat{\section}{\normalsize\bfseries}{\thesection.}{0.5em}{}
\titlespacing*{\section}{0pt}{2.5ex plus 0.2ex minus 0.2ex}{1.0ex}
\titleformat{\subsection}{\normalsize\itshape}{\thesubsection}{0.5em}{}
\titlespacing*{\subsection}{0pt}{1.5ex plus 0.2ex minus 0.2ex}{0.8ex}
\titleformat{\subsubsection}{\normalsize\itshape}{\thesubsubsection}{0.6em}{}
\titlespacing*{\subsubsection}{0pt}{1.2ex plus 0.2ex minus 0.2ex}{0.6ex}
\titleformat{\paragraph}{\normalsize\bfseries}{}{0em}{}
\titlespacing*{\paragraph}{0pt}{1.2ex}{0.2em}

\usepackage[font=small,labelfont=bf]{caption}
\DeclareMathOperator{\Var}{Var}

\DeclareMathOperator{\CV}{CV}

\DeclareMathOperator{\PPM}{PPM}
\DeclareMathOperator{\LCB}{LCB}

\newcommand{\USL}{\mathrm{USL}}
\newcommand{\LSL}{\mathrm{LSL}}

\newcommand{\Tol}{\mathrm{Tol}}

\usepackage{threeparttable}
\usepackage{algorithm}
\usepackage{algpseudocode}

\usepackage[dvipsnames]{xcolor}
\usepackage[normalem]{ulem}
\usepackage{soul}
\sethlcolor{yellow!35}

\usepackage{framed}

\title{Nonlinear Amplification of Finite-Sample Uncertainty in Capability-Based Decisions}

\author{
	Fei Jiang\thanks{Independent researchers, Seattle, WA, USA.
		Corresponding author: Fei Jiang, email: jiangfeicq@gmail.com.}
	\and
	Lei Yang\footnotemark[1]
}

\date{}
\begin{document}
\maketitle

\begin{abstract}
	This paper studies the propagation of finite-sample uncertainty under nonlinear transformations commonly used in statistical decision systems. In particular, we consider process capability indices, which are widely used in manufacturing practice but are estimated from finite samples, rendering the resulting approval decisions inherently uncertain. We show that such uncertainty cannot be fully explained by estimator variability alone, but is substantially influenced by a nonlinear amplification mechanism through which capability uncertainty is transformed into defect-risk metrics. While capability estimators vary approximately linearly with process dispersion, defect probabilities depend on tail curvature, causing small estimation errors to be disproportionately amplified in measures such as defect probability and parts-per-million (PPM) rates. Consequently, capability assessments that appear stable in index space may exhibit substantial variability in defect-risk space, particularly near decision thresholds. This insight provides a unified explanation of finite-sample decision instability, motivates reliability-aware decision formulations, and links sample-size requirements directly to decision reliability. Monte Carlo simulations and industrial data analyses validate the proposed mechanism and demonstrate its practical implications, including the impact of distributional assumptions on defect-risk estimation.
\end{abstract}


%\tableofcontents
%%%%%%%%%%%%%%%%%%%%%%%%%%%%%%%%%%%%%%%%%%%
%\newpage
\section{Introduction}
\label{sec:introduction}

Process capability indices (PCIs) are fundamental tools in statistical quality engineering for assessing whether a manufacturing process can consistently meet specification limits \cite{kane1986process,anis2008basic,kotz2002process,montgomery2020introduction,wu2009overview}. Classical indices such as $C_p$, $C_{pk}$, and their variants quantify the relationship between process variation and tolerance bounds and are widely used in supplier qualification, process validation, and production release decisions under standardized frameworks such as ISO/TR~22514-4 \cite{ISO22514-4-2016,ISO11462-1-2001,ISO22514-1-2014}. In practice, capability thresholds (e.g., $C_{pk} \ge 1.33$) are commonly interpreted as indicators of acceptable defect risk and process reliability.

A substantial amount of research has focused on statistical inference for capability indices, including point estimation, confidence intervals, and hypothesis testing, as well as extensions to non-normal processes and alternative estimators \cite{pearn1992distributional,kushler1992confidence,mathew2007generalized,zhang1990interval,mahmoud2010estimating,alvarez2015methodological,chen1997application,chen2001new,clements1989process,collins1995bootstrap}. While these studies provide a comprehensive framework for capability estimation, they generally treat capability indices as stable summaries of process quality and do not explicitly address how estimation uncertainty propagates into downstream decision making.

Because capability indices are computed from finite samples, the estimator $\widehat{C}_{pk}$ is inherently random. Prior work has shown that threshold-based capability approval should be interpreted as a statistical decision problem rather than a deterministic rule, particularly when the true capability lies near the acceptance boundary \cite{jiang2026finite,bissell1990reliable}. In such regimes, repeated samples from the same process may lead to inconsistent approval outcomes. Related developments have proposed structured workflows and risk-calibrated decision rules that explicitly incorporate estimation uncertainty into capability-based decisions \cite{jiang2026practical,jiang2026risk,pendrill2014using}.

Despite these advances, a fundamental question remains insufficiently understood: how estimation uncertainty in capability indices propagates into defect-risk metrics that underpin engineering decisions. In practice, capability indices are often interpreted as proxies for defect probability or parts-per-million rates \cite{bebr2017use,deleryd1998gap}. However, these quantities are linked through inherently nonlinear transformations, and statistical decision theory suggests that such mappings can substantially distort uncertainty propagation \cite{degroot2005optimal,lehmann2005testing,berger2013statistical}. While this effect is well recognized in general statistical settings, its role in capability-based decision systems has not been explicitly characterized. In particular, existing capability analysis has largely focused on estimator properties and threshold rules, rather than on how uncertainty is structurally amplified when mapped into defect-risk space. This work shows that such nonlinear effects play a central role in governing finite-sample decision instability.

This paper demonstrates that the instability of capability-based decisions under finite samples is governed by a nonlinear amplification mechanism through which estimation uncertainty in capability indices is transformed into defect-risk metrics. For the normal case, this mechanism is further quantified through a closed-form amplification coefficient that measures the relative sensitivity of defect-risk metrics to dispersion perturbations. While capability estimators vary approximately linearly with process dispersion, defect probabilities are determined by the curvature of distribution tails, causing small estimation errors to be disproportionately amplified when translated into reliability measures such as defect probability or parts-per-million rates. As a result, capability assessments that appear stable in index space may exhibit substantial variability in defect-risk space, particularly near decision thresholds.

This paper makes three main contributions. First, we identify a nonlinear amplification mechanism through which estimation uncertainty in capability indices is transformed into defect-risk variability, revealing that variability which appears moderate in capability space can be structurally amplified in tail-risk metrics due to distributional curvature. Second, we establish a quantitative link between capability indices and defect-risk sensitivity by deriving an amplification coefficient, providing a direct connection between process capability level and reliability sensitivity. Third, we show that this mechanism fundamentally reshapes capability-based decision behavior, leading to elevated misclassification risk near thresholds and motivating reliability-based decision rules and sample-size criteria.

The remainder of this paper is organized as follows. 
Section~\ref{sec:capability_estimation} introduces notation and background on capability estimation. 
Section~\ref{sec:nonlinear_amplification} presents the nonlinear amplification analysis. 
Section~\ref{sec:decision_framework} discusses implications for capability-based decision making. 
Section~\ref{sec:sample_size} examines sample-size considerations. 
Section~\ref{sec:empirical} provides empirical validation, and Section~\ref{sec:conclusion} concludes the paper.

%%%%%%%%%%%%%%%%%%%%%%%%%%%%%%%%%%%%%%%
\section{Background: Capability Estimation}
\label{sec:capability_estimation}

The process capability index $C_{pk}$ is one of the most commonly used metrics in industrial capability assessment \cite{montgomery2020introduction}.

For clarity, we focus on normally distributed processes with bilateral specification limits, which provide a tractable setting for developing the amplification mechanism. Extensions beyond normality are discussed later.

Consider a quality characteristic $X$ with lower and upper specification limits $\mathrm{LSL}$ and $\mathrm{USL}$, and process parameters $(\mu,\sigma)$. The PCI $C_{pk}$ is defined as
\[
C_{pk}
=
\min
\left(
\frac{\mathrm{USL}-\mu}{3\sigma},
\frac{\mu-\mathrm{LSL}}{3\sigma}
\right).
\]

In practice, the process parameters $(\mu,\sigma)$ are unknown and must be estimated from a finite sample $\{X_1,\dots,X_n\}$ of size $n$. Let
\[
\bar{X}=\frac{1}{n}\sum_{i=1}^{n}X_i,
\qquad
S^2=\frac{1}{n-1}\sum_{i=1}^{n}(X_i-\bar{X})^2
\]
denote the sample mean and sample variance. The standard plug-in estimator of process capability is
\[
\widehat{C}_{pk}
=
\min
\left(
\frac{\mathrm{USL}-\bar{X}}{3S},
\frac{\bar{X}-\mathrm{LSL}}{3S}
\right).
\]

Under normal sampling, the variability of the sample standard deviation $S$ can be quantified through its coefficient of variation. Using the chi-square distribution of $(n-1)S^2/\sigma^2$, a standard approximation gives \cite{montgomery2020introduction}
\begin{equation}
	\CV(S)
	\approx
	\frac{1}{\sqrt{2(n-1)}},
	\label{eq:cvS_approx}
\end{equation}
indicating that dispersion estimation remains non-negligibly variable even at moderate sample sizes.

Because $\bar{X}$ and $S$ are random, $\widehat{C}_{pk}$ is stochastic under finite samples. Consequently, approval rules of the form
\begin{equation}
	\widehat{C}_{pk} \ge C_0
	\label{eq:cpk_rule}
\end{equation}
involve statistical decision uncertainty, with small fluctuations near $C_0$ potentially leading to different outcomes.

The statistical properties of $\widehat{C}_{pk}$, including its finite-sample variability and the resulting decision instability, have been analyzed in \cite{jiang2026practical}, which shows that threshold-based classification may exhibit substantial misclassification risk under limited sample sizes. Building on these results, the present paper investigates how estimation uncertainty propagates from capability index space into defect-risk metrics such as defect probability and PPM rates. As shown in subsequent sections, this propagation is governed by a nonlinear amplification mechanism, under which moderate variability in $\widehat{C}_{pk}$ can be substantially magnified in defect-risk space.

%%%%%%%%%%%%%%%%%%%%%%%%%%%%%%%%%%%%%%%%%%%%%%
\section{Nonlinear Amplification of Capability Uncertainty}
\label{sec:nonlinear_amplification}

While prior work has established the asymptotic boundary instability of capability classification \cite{jiang2026finite}, this section identifies the structural mechanism underlying decision sensitivity under finite samples. Specifically, we show that estimation uncertainty is nonlinearly amplified when capability indices are interpreted in terms of defect-risk metrics.

Although capability approval rules are defined in terms of $\widehat{C}_{pk}$, capability indices are commonly used as surrogates for defect probability. This induces a nonlinear mapping from estimation space to tail-risk space, under which small errors in $(\bar{X}, S)$ may be substantially magnified when translated into reliability metrics. As a result, variability that appears moderate in capability space can produce large changes in defect risk, leading to elevated decision sensitivity near the capability threshold.

% =========================================================
\subsection{Mapping Capability to Defect Probability}
\label{sec:nonlinear_amplification-a}

Capability indices are commonly interpreted in terms of defect probability, linking higher capability to lower risk of exceeding specification limits. Let
\begin{equation}
	p_{\text{defect}} = \Pr(X \notin [\LSL,\USL]), 
	\PPM = 10^6\times p_{\text{defect}}.
	\label{eq:defect_def}
\end{equation}

Defect probability is determined by the tail behavior of the process distribution. For a distribution with cumulative distribution function $F$,
\[
p_{\text{defect}}
=
F\!\left(\frac{\LSL-\mu}{\sigma}\right)
+
1 - F\!\left(\frac{\USL-\mu}{\sigma}\right).
\]
Under normality, this reduces to the Gaussian expression with $F=\Phi$.

This formulation highlights a nonlinear mapping from $(\mu,\sigma)$ to defect risk, under which small perturbations in process parameters can be substantially amplified when translated into reliability metrics.

% =========================================================
\subsection{Nonlinear Amplification Mechanism}

Let $\widehat{\PPM}=10^6\,h(\bar{X},S)$ denote the estimated defect rate obtained by replacing $(\mu,\sigma)$ with their sample estimators. The mapping $h(\cdot)$ captures the transformation from process parameters to defect probability and is governed by tail behavior of the underlying distribution.

To characterize sensitivity, consider the upper-tail defect probability
\begin{equation}
	p_{\text{upper}} = 1 - F(z), 
	\qquad 
	z = \frac{\USL-\mu}{\sigma}.
\end{equation}
Differentiating with respect to $\sigma$ yields
\begin{equation}
	\frac{\partial p_{\text{upper}}}{\partial \sigma}
	=
	f(z)\,\frac{\USL-\mu}{\sigma^2},
\end{equation}
where $f$ is the density associated with $F$. This shows that dispersion uncertainty is scaled by the tail density at the standardized boundary. Because $f(z)$ decays nonlinearly with $z$, small perturbations in $\sigma$ can induce disproportionately large changes in defect probability in tail-dominated regimes.

To quantify this effect, we define the amplification coefficient
\begin{equation}
	A_{\sigma}
	=
	\left|
	\frac{\partial \log \PPM}{\partial \log \sigma}
	\right|,
	\label{eq:amp_sigma_def}
\end{equation}
which measures the sensitivity of defect-risk metrics to relative dispersion perturbations. Values of $A_{\sigma}>1$ indicate amplification.

From this definition, $A_{\sigma}$ increases with the standardized distance to the specification limits, implying a monotone relationship with $C_{pk}$ and stronger amplification at higher capability levels due to tail curvature effects.

For a centered symmetric process under normality, let
\[
z=\frac{\USL-\mu}{\sigma}=\frac{\mu-\LSL}{\sigma}>0.
\]
Then
\[
p_{\text{defect}}=2\{1-\Phi(z)\},
\qquad
\PPM = 2\times 10^6 \{1-\Phi(z)\},
\]
and direct differentiation yields
\begin{equation}
	A_{\sigma}
	=
	\frac{z\phi(z)}{1-\Phi(z)}
	=
	z\,r(z),
	\label{eq:amp_sigma_normal}
\end{equation}
where $r(z)=\phi(z)/(1-\Phi(z))$ is the Mills ratio, which characterizes the curvature of the distribution tail. Here $z = (USL - \mu)/\sigma$ denotes the standardized distance from the process mean to the specification limit. Under the centered case, this reduces to $z = 3C_{pk}$, linking the capability index to the tail location.

Substituting this relation yields
\begin{equation}
	A_{\sigma}
	=
	3C_{pk}\,r(3C_{pk}),
	\label{eq:amp_sigma_cpk}
\end{equation}
which directly connects the capability level to amplification strength through tail curvature.

This result highlights that amplification is governed by the interaction between the standardized distance to the specification limit and tail curvature: as processes enter tail-dominated regimes, small relative perturbations in dispersion induce large changes in defect probability.

Using a first-order approximation via the standard Delta method \cite{van2000asymptotic}, the variance of $\widehat{\mathrm{PPM}}$ can be approximated as
\begin{equation}
	\Var(\widehat{\PPM})
	\approx
	\nabla h^\top
	\Sigma_{\bar{X}, S}
	\nabla h,
\end{equation}
where $\Sigma_{\bar{X}, S}$ is the covariance of $(\bar{X}, S)$. 

In contrast, capability indices depend approximately linearly on $\sigma^{-1}$, leading to substantially lower variability relative to defect-risk metrics. Consequently,
\begin{equation}
	\Var(\widehat{\PPM}) \gg \Var(\widehat{C}_{pk}),
\end{equation}
particularly near common capability thresholds.

This inequality can be further understood through the Delta method. In particular, the variance of $\widehat{\PPM}$ is scaled by the squared gradient of the mapping from $(\bar{X},S)$ to defect probability, which increases rapidly in tail regions. This effect is captured by the amplification coefficient $A_{\sigma}$, showing that variance propagation in defect-risk space is intrinsically magnified relative to capability space.

This disparity reveals a structural amplification effect: estimation uncertainty that appears moderate in capability space becomes significantly magnified when mapped into defect-risk space. Near the classification boundary $C_0$, this effect translates symmetric estimator variability into asymmetric decision risk, concentrating misclassification probability in a narrow region.

Although the derivation assumes normality for analytical transparency, the mechanism is not distribution-specific. 
This effect is further examined empirically in Section~\ref{sec:empirical}, where defect-risk estimates under fitted distributions are compared with normal-based approximations.
More generally, the amplification coefficient can be expressed as
\[
A_{\sigma} = \left| \frac{\partial \log p_{\text{tail}}}{\partial \log \sigma} \right|,
\]
where $p_{\text{tail}}$ denotes the relevant tail probability. This shows that amplification is governed by local tail behavior and can be interpreted through survival or hazard functions. For skewed or heavy-tailed distributions, slower tail decay increases sensitivity, implying stronger amplification than in the normal case.

While the present analysis focuses on dispersion-driven amplification, mean uncertainty may also contribute to defect-risk sensitivity in off-center processes. Extending the framework to jointly account for location and scale perturbations is a direction for future work.

%%%%%%%%%%%%%%%%%%%%%%%%%%%%%%%%%%%%%%%%%%%%%%%%%%%%%%%%%%%%%%%%%%%%
\section{Reliability-Aware Decision Framework}
\label{sec:decision_framework}

Building on the amplification mechanism established in Section~\ref{sec:nonlinear_amplification}, we now focus on its implications for capability-based decision making. In practice, capability approval is implemented through threshold rules applied to $\widehat{C}_{pk}$, which is random under finite samples. This motivates a decision-theoretic formulation that explicitly characterizes misclassification risk and incorporates estimation uncertainty into approval rules. This stochasticity is further amplified when mapped into defect-risk space, increasing decision sensitivity near the threshold.

\subsection{Misclassification Risk}

Let $C_{pk}^{true}$ denote the true population capability. Misclassification occurs when the decision based on $\widehat{C}_{pk}$ disagrees with the true capability status relative to the threshold $C_0$. This can be characterized by the Type I and Type II error probabilities
\[
\begin{aligned}
	P_{\text{Type I}} &= \Pr(\widehat{C}_{pk} \ge C_0 \mid C_{pk}^{\text{true}} < C_0), \\
	P_{\text{Type II}} &= \Pr(\widehat{C}_{pk} < C_0 \mid C_{pk}^{\text{true}} \ge C_0).
\end{aligned}
\]

While these probabilities arise from finite-sample variability in $\widehat{C}_{pk}$, they are particularly sensitive to estimation fluctuations near the threshold, where small changes in $\widehat{C}_{pk}$ can lead to different classification outcomes.

\subsection{Reliability-Based Capability Approval}

To mitigate amplification-induced instability, capability approval can be formulated as a probabilistic decision problem requiring
\[
\Pr(C_{pk}^{true} \ge C_0 \mid \text{data}) \ge \gamma,
\]
where $\gamma$ is a prescribed reliability level.

In frequentist practice, this is typically implemented via a one-sided lower confidence bound (LCB):
\begin{equation}
	\mathrm{LCB}_{1-\alpha}(\widehat{C}_{pk}) \ge C_0,
	\quad \alpha = 1 - \gamma.
	\label{eq:lcb_rule}
\end{equation}

Compared with deterministic thresholding, this formulation explicitly incorporates estimation uncertainty into the decision rule. From the perspective of nonlinear amplification, it controls the propagation of estimator variability into decision outcomes, thereby reducing sensitivity near the threshold.

\subsection{Implications for Sample Size Planning}

Because amplification magnifies estimator variability near the capability threshold, decision reliability depends critically on sample size. Although the variance of $\widehat{C}_{pk}$ decreases with increasing $n$, the nonlinear mapping to defect-risk space implies that moderate estimation uncertainty may still produce substantial decision variability in small-sample regimes.

This observation reveals that sample size plays a fundamentally different role from classical variance reduction: it governs the extent to which nonlinear amplification translates estimation uncertainty into decision instability. Consequently, sample size requirements should be formulated in terms of decision reliability rather than estimator precision alone.

This perspective provides a mechanism-based foundation for capability study design and motivates a formal characterization of decision reliability as a function of sample size, developed in the following section.

%%%%%%%%%%%%%%%%%%%%%%%%%%%%%%%%%%%%%%%%%%%%%%%%
% =====================================================================
%%%%%%%%%%%%%%%%%%%%%%%%%%%%%%%%%%%%%%%%%%%%%%%%
\section{Sample Size for Controlling Amplification-Induced Misclassification}
\label{sec:sample_size}

This section examines how sample size affects decision reliability in capability-based approval. Under finite samples, classification outcomes based on $\widehat{C}_{pk}$ depend on sampling variability, particularly near the threshold $C_0$. This motivates a stochastic decision formulation in which sample size governs how estimation uncertainty translates into misclassification risk under nonlinear amplification. Unlike classical approaches that focus on estimator variance or confidence interval width, the present framework links sample size directly to the probability of correct classification. In what follows, we formalize this relationship through acceptance probability and derive conditions for controlling misclassification risk.

\subsection{Misclassification Risk and Acceptance Probability}

Let $C_{pk}^{true}$ denote the true population capability, and let $n$ denote the sample size used to estimate $\widehat{C}_{pk}$ from independent observations. Misclassification occurs when the decision based on $\widehat{C}_{pk}$ disagrees with the true capability status relative to the threshold. This can be characterized through the acceptance probability
\begin{equation}
	p_{\mathrm{acc}}(n)
	= \Pr\!\left(\widehat{C}_{pk,n} \ge C_0 \mid C_{pk}^{true}\right),
	\label{eq:pacc_def_refined}
\end{equation}
which represents the probability that the process is classified as capable under repeated sampling of size $n$.

When $C_{pk}^{true} < C_0$, $p_{\mathrm{acc}}(n)$ corresponds to false-acceptance risk, while for $C_{pk}^{true} \ge C_0$, $1 - p_{\mathrm{acc}}(n)$ represents false-rejection risk. In particular, near the threshold $C_0$, $p_{\mathrm{acc}}(n)$ may vary sharply with $n$, reflecting the sensitivity of classification outcomes to finite-sample variability.

The bootstrap approximation provides a practical way to estimate $p_{\mathrm{acc}}(n)$ under limited distributional assumptions by approximating repeated sampling from the empirical distribution. While it does not recover the true sampling distribution, it offers a data-driven estimate of decision reliability across different sample sizes.

\subsection{Reliability-Based Sample Size Definition}

To control such instability, the minimum sample size can be defined in terms of decision reliability. Given a target reliability level $\gamma$, define
\begin{equation}
	n_{\min}(\gamma; C_{pk}^{true})
	=
	\min\left\{ n : p_{\mathrm{acc}}(n) \ge \gamma \right\}.
	\label{eq:nmin_refined}
\end{equation}

This formulation characterizes sample size as a decision-theoretic quantity, linking statistical uncertainty directly to misclassification risk.

From the perspective of nonlinear amplification, $n_{\min}$ represents the minimum sample size required to suppress amplification-induced variability to an acceptable level. In contrast to classical approaches that focus solely on estimator precision, this formulation explicitly incorporates the amplification effect arising from the nonlinear mapping to defect-risk space.

An equivalent conservative formulation is based on lower confidence bounds (LCB), as given in Eq.~\eqref{eq:lcb_rule}. Such rules explicitly incorporate estimator uncertainty and can be interpreted as controlling the propagation of amplified variability into decision outcomes.

\subsection{Implications for Capability Study Design}

\begin{figure}[!t]
	\centerline{\includegraphics[width=1.00\linewidth]{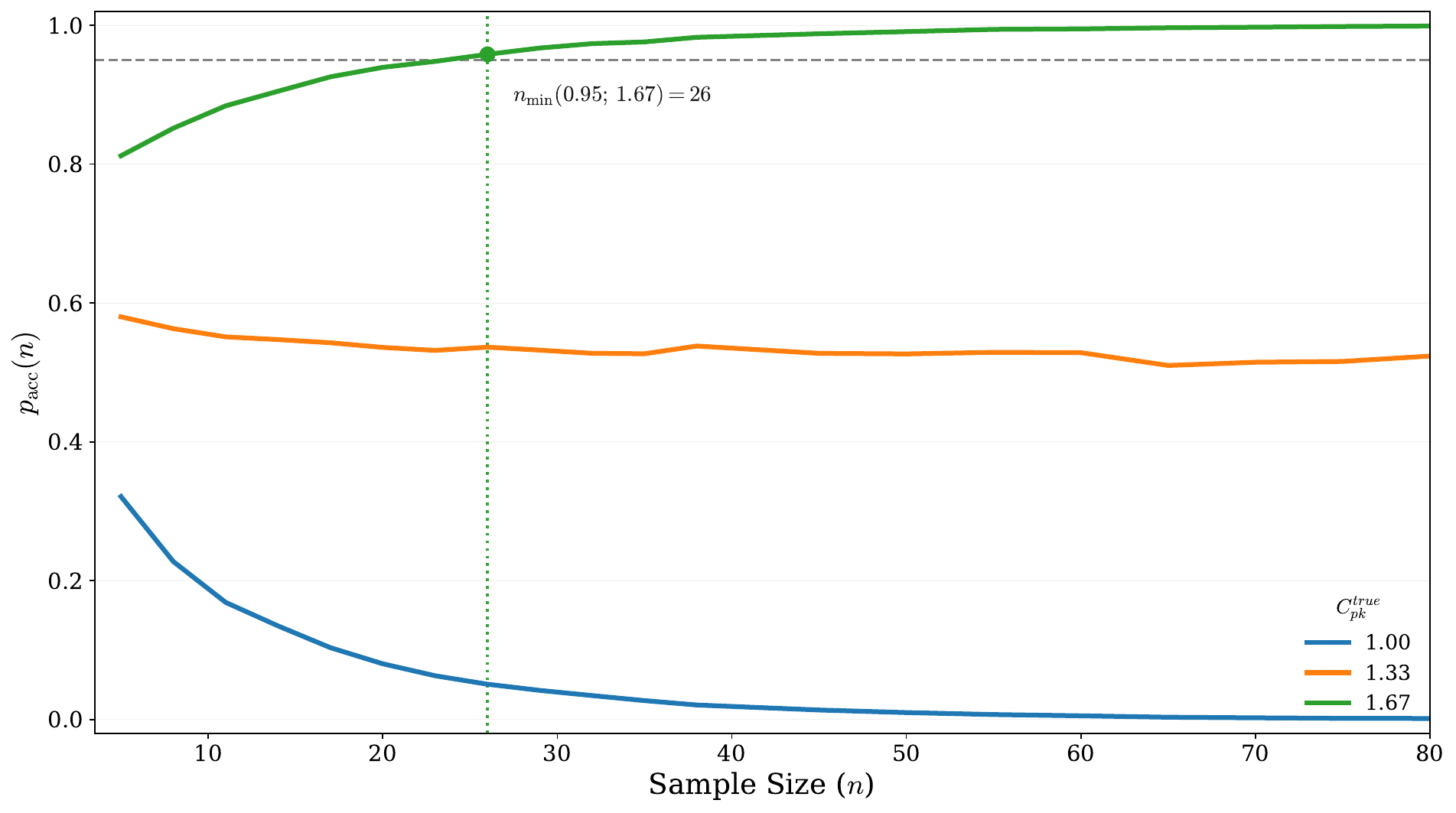}}
	\caption{
		Monte Carlo evaluation of the probability-of-pass function 
		$p_{\mathrm{acc}}(n)=\Pr(\widehat{C}_{pk,n}\ge C_0 \mid C_{pk}^{true})$
		for $C_0=1.33$ under centered normal sampling with symmetric bilateral specifications. 
		Curves correspond to $C_{pk}^{true}\in\{1.00,\,1.33,\,1.67\}$, and the dashed horizontal line indicates the reliability target $\gamma=0.95$. Each curve is estimated using $R=5{,}000$ replications per sample size. The figure shows a sharp transition in acceptance probability near the threshold, indicating high sensitivity of classification outcomes to finite-sample variability. At $C_{pk}^{true}=C_0$, the acceptance probability is close to $0.5$, with minor deviations attributable to finite-sample effects.
	}
	\label{fig:prob_pass}
\end{figure}

To construct Figure~\ref{fig:prob_pass}, we consider centered normal processes with symmetric bilateral specifications and target values of $C_{pk}^{true}\in\{1.00,1.33,1.67\}$. For each combination of $n$ and $C_{pk}^{true}$, independent samples are generated from the corresponding normal model, the plug-in estimator $\widehat{C}_{pk}$ is computed, and the deterministic approval rule $\widehat{C}_{pk}\ge C_0$ is evaluated. The acceptance probability $p_{\mathrm{acc}}(n)$ is then estimated as the proportion of accepted samples over repeated Monte Carlo replications. The simulation setup isolates finite-sample variability by fixing the process mean and symmetric specification limits.

Figure~\ref{fig:prob_pass} shows how acceptance probability evolves with sample size under different values of $C_{pk}^{true}$. When $C_{pk}^{true}$ lies well below or above the threshold, $p_{\mathrm{acc}}(n)$ varies smoothly with $n$, indicating limited sensitivity to estimator variability. In contrast, when $C_{pk}^{true}$ is close to $C_0$, the acceptance probability exhibits a sharp transition, where small fluctuations in $\widehat{C}_{pk}$ produce large changes in classification outcomes.

These results imply that sample size should be determined based on desired decision reliability rather than estimator precision alone. In particular, near-threshold processes require substantially larger sample sizes to achieve stable classification. For non-normal or tail-sensitive processes, data-driven evaluation of $p_{\mathrm{acc}}(n)$ provides a practical approach to determining $n_{\min}$ under realistic conditions.

%%%%%%%%%%%%%%%%%%%%%%%%%%%%%%%%%%%%%%%%%%%%%
\section{Empirical Validation Using Large-Scale Manufacturing Data}
\label{sec:empirical}

This section empirically validates the theoretical implications of the proposed amplification framework using large-scale manufacturing data. 

The theoretical development is based on $C_{pk}$ under controlled process assumptions. In the empirical analysis, capability values are computed using data-driven approaches that may involve overall variation or distribution-aware formulations for non-normal processes. We retain the notation $C_{pk}$ for consistency, as the amplification mechanism depends on standardized distance to specification limits and tail behavior rather than the specific estimation method.

Using a 500-dimension manufacturing dataset, from which 18 dimensions were selected using a stratified procedure to ensure representative coverage of different capability levels and distances to the decision threshold, the empirical study is organized into three parts. First, we examine the sensitivity of reported capability to distributional assumptions under fixed sample size and symmetric specifications. Second, we evaluate how finite-sample variability is reflected in defect-risk estimates. Third, we analyze the prevalence of near-threshold dimensions and their impact on decision stability. Together, these results provide empirical evidence on how estimation uncertainty propagates into capability-based decisions in practice.

%%%%%%%%%%%%%%%%%%%%%%%%%%%%%%%%%%%%%%%%%%%%%%%%%%
\subsection{Data Description and Evaluation Framework}
\label{sec:val_setup}

We analyze a manufacturing dataset in which each dimension is measured $n=32$ times, with nominal values $T_j$ and symmetric bilateral tolerances defining specification limits $\USL_j = T_j + \Tol_j$ and $\LSL_j = T_j - \Tol_j$. For data protection, dimension-wise constant offsets are applied to both measurements and specification limits; these affine shifts preserve tolerance widths and do not affect capability or decision-related quantities.

For each dimension, we report both the normal-assumption benchmark $\widehat{C}_{pk}^{\mathrm{Normal}}$ and a distribution-aware estimate $\widehat{C}_{pk}$. When normality is rejected, the latter is obtained from the best-fitting parametric model selected from Normal, Lognormal, Weibull, and Logistic families using the small-sample corrected Akaike Information Criterion (AICc), which is appropriate for finite-sample settings. This procedure provides a systematic assessment of sensitivity to distributional assumptions.

Decision reliability is evaluated via nonparametric bootstrap resampling. For each dimension, $B=2000$ bootstrap samples of size $n=32$ are drawn with replacement, and the approval indicator $\mathbb{I}(\widehat{C}_{pk}\ge C_0)$ is recorded. The bootstrap pass probability $\widehat{p}$ is defined as the fraction of accepted resamples, and $\LCB_{0.95}$ denotes the corresponding one-sided 95\% lower confidence bound.

To align with the theoretical setting, we restrict attention to dimensions with symmetric two-sided specifications and nonnegative capability estimates. From the eligible pool, 18 dimensions are selected using a stratified procedure based on estimated capability levels, distance to the decision threshold, and distributional characteristics, ensuring coverage of low-, mid-, and high-capability regimes, near-threshold cases, and both approximately normal and non-normal distributions. This strategy reduces selection bias and improves representativeness relative to the full dataset. Summary statistics and defect-rate comparisons are reported in Table~\ref{tab:val_stats18}, and raw measurements are provided in Appendix~\ref{app:raw-data}.

\begin{table*}[htbp]
	\centering
	\footnotesize
	\renewcommand{\arraystretch}{1.0}
	\setlength{\tabcolsep}{2.0pt}
	\resizebox{\textwidth}{!}{%
		\begin{tabular}{cccccccccccccccccccc}
			\toprule
%			& & & & & & & & 
			& & & & 
			& \multicolumn{4}{c}{Diagnostics}
			& \multicolumn{3}{c}{Capability}
			& \multicolumn{2}{c}{Decision}
			& \multicolumn{3}{c}{Fitted PPM}
			& \multicolumn{3}{c}{Normal PPM} \\
			\cmidrule(lr){6-9}
			\cmidrule(lr){10-12}
			\cmidrule(lr){13-14}
			\cmidrule(lr){15-17} \cmidrule(lr){18-20}
			Dim & $T$ & Tol$\pm$ & $\bar{X}$ & S & Skew & Kurt & P.value & Dist
			& $\widehat{C}_{pk}$ & $\widehat{C}_{pk}^{\mathrm{Normal}}$ & $\Delta C_{pk}$
			& $\widehat{p}$ & LCB$_{0.95}$
			& $\widehat{\PPM}_0$ & $\widehat{\PPM}_+$ & $\widehat{\PPM}_-$
			& $\widehat{\PPM}_0^{\mathrm{N}}$ & $\widehat{\PPM}_+^{\mathrm{N}}$ & $\widehat{\PPM}_-^{\mathrm{N}}$ \\
			\midrule
			D056 & 0.55 & 0.10 & 0.519 & 0.0172 & -1.352 & 1.280 & 0.000 & Weibull & 1.071 & 1.332 & -0.261 & 0.543 & 1.033 & 874 & 1813 & 341 & 32.1 & 195 & 2.34 \\
			D080 & 1.17 & 0.10 & 1.178 & 0.0230 & 1.068 & 0.953 & 0.036 & Logistic & 1.148 & 1.339 & -0.191 & 0.558 & 1.053 & 732 & 1756 & 238 & 31.0 & 199 & 2.15 \\
			D090 & 1.65 & 0.05 & 1.646 & 0.0116 & -0.099 & -0.323 & 0.565 & Normal & 1.334 & 1.334 & 0.000 & 0.575 & 1.132 & 33.1 & 211 & 2.32 & 33.1 & 211 & 2.32 \\
			D117 & 2.27 & 0.10 & 2.278 & 0.0231 & 0.023 & -0.278 & 0.064 & Normal & 1.334 & 1.334 & 0.000 & 0.576 & 1.133 & 32.9 & 209 & 2.33 & 32.9 & 209 & 2.33 \\
			D158 & 2.74 & 0.10 & 2.724 & 0.0214 & -0.074 & -1.138 & 0.202 & Normal & 1.304 & 1.304 & 0.000 & 0.479 & 1.146 & 45.7 & 259 & 3.70 & 45.7 & 259 & 3.70 \\
			D186 & 3.28 & 0.10 & 3.291 & 0.0228 & -1.106 & 0.525 & 0.002 & Weibull & 1.151 & 1.308 & -0.157 & 0.510 & 1.149 & 485 & 1069 & 175 & 44.1 & 257 & 3.49 \\
			D224 & 3.70 & 0.05 & 3.697 & 0.0121 & 0.266 & -1.130 & 0.042 & Lognormal & 1.338 & 1.309 & 0.029 & 0.486 & 1.168 & 35.4 & 233 & 2.37 & 50.0 & 304 & 3.76 \\
			D227 & 4.21 & 0.10 & 4.238 & 0.0185 & 1.950 & 6.115 & 0.000 & Logistic & 1.568 & 1.303 & 0.265 & 0.523 & 0.921 & 31.7 & 98.1 & 7.39 & 46.5 & 263 & 3.79 \\
			D228 & 5.10 & 0.05 & 5.087 & 0.0090 & 0.802 & 0.708 & 0.016 & Logistic & 1.129 & 1.379 & -0.250 & 0.693 & 1.173 & 580 & 1373 & 191 & 17.7 & 121 & 1.08 \\
			D252 & 5.52 & 0.25 & 5.578 & 0.0468 & -0.343 & -0.339 & 0.535 & Normal & 1.370 & 1.370 & 0.000 & 0.683 & 1.184 & 19.8 & 133 & 1.25 & 19.8 & 133 & 1.25 \\
			D253 & 6.16 & 0.05 & 6.123 & 0.0034 & 0.428 & -1.258 & 0.002 & Lognormal & 1.308 & 1.287 & 0.021 & 0.411 & 1.175 & 42.8 & 246 & 3.41 & 56.3 & 305 & 4.85 \\
			D288 & 6.65 & 0.05 & 6.624 & 0.0060 & 1.145 & 1.169 & 0.008 & Logistic & 1.106 & 1.367 & -0.261 & 0.667 & 1.165 & 639 & 1499 & 212 & 20.5 & 137 & 1.31 \\
			D308 & 7.13 & 0.10 & 7.154 & 0.0074 & 0.459 & 0.258 & 0.255 & Normal & 3.420 & 3.420 & 0.000 & 1.000 & 2.875 & 9.24e-58 & 1.19e-44 & 5.50e-77 & 5.38e-19 & 4.39e-14 & 3.47e-26 \\
			D318 & 7.78 & 0.05 & 7.778 & 0.0072 & -1.585 & 4.057 & 0.009 & Logistic & 2.001 & 2.228 & -0.227 & 1.000 & 1.756 & 2.18 & 10.2 & 0.301 & 1.18e-05 & 1.57e-03 & 9.65e-09 \\
			D360 & 8.20 & 0.10 & 8.237 & 0.0487 & -0.584 & -0.281 & 0.089 & Normal & 0.431 & 0.431 & 0.000 & 0.000 & 0.355 & 1.00e5 & 1.32e5 & 6.99e4 & 1.00e5 & 1.32e5 & 6.99e4 \\
			D390 & 8.65 & 0.10 & 8.720 & 0.0220 & -1.061 & 0.389 & 0.002 & Weibull & 0.753 & 0.455 & 0.298 & 0.000 & 0.394 & 2.29e4 & 4.82e4 & 6.61e3 & 8.61e4 & 1.13e5 & 5.89e4 \\
			D401 & 9.19 & 0.05 & 9.191 & 0.0087 & -0.763 & -0.100 & 0.006 & Weibull & 1.241 & 1.886 & -0.645 & 1.000 & 1.597 & 295 & 689 & 98.3 & 1.08e-02 & 3.86e-01 & 5.99e-05 \\
			D410 & 9.87 & 0.08 & 9.845 & 0.0132 & -0.151 & -0.780 & 0.709 & Normal & 1.379 & 1.379 & 0.000 & 0.692 & 1.188 & 17.7 & 121 & 1.08 & 17.7 & 121 & 1.08 \\
			\bottomrule
		\end{tabular}%
	}
	\caption{Integrated validation results for the 18 selected dimensions from the 500-dimension dataset. The table combines distribution diagnostics (Skew, Kurt, P.value, Dist), model-induced capability distortion ($\Delta C_{pk}=\widehat{C}_{pk}-\widehat{C}_{pk}^\mathrm{Normal}$), bootstrap decision reliability ($\widehat{p}$ and $\LCB_{0.95}$), and nonlinear defect-rate amplification. Defect rates are evaluated under the normal assumption as $\widehat{\PPM}_0=\widehat{\PPM}(S)$, $\widehat{\PPM}_{+}=\widehat{\PPM}(S(1+\CV))$, and $\widehat{\PPM}_{-}=\widehat{\PPM}(S(1-\CV))$, where $\CV$ is given by Eq.~\eqref{eq:cvS_approx}. For $n=32$, $\CV \approx 0.127$.}
	\label{tab:val_stats18}
\end{table*}

%%%%%%%%%%%%%%%%%%%%%%%%%%%%%%%%%%%%%%%%%%%%%%%%%%
\subsection{Model-Induced Amplification Sensitivity}
\label{sec:val_sec2}

While the preceding analysis focuses on finite-sample variability under a fixed model, amplification effects can also arise from model uncertainty. In practice, capability assessment depends on the assumed data-generating distribution, and small differences in model choice may be amplified when translated into defect-risk metrics.

To quantify this effect, we define the model-induced distortion
\begin{equation}
	\Delta C_{pk,j}
	=
	\widehat{C}_{pk,j}
	-
	\widehat{C}_{pk,j}^{\mathrm{Normal}},
	\label{eq:delta_model_val}
\end{equation}
where $\widehat{C}_{pk,j}$ is computed under the selected parametric model for dimension $j$, and $\widehat{C}_{pk,j}^{\mathrm{Normal}}$ is computed under the normality assumption. This quantity captures model-induced variation in capability space, which serves as the input to amplification in defect-risk space.

Across the 18 selected dimensions, $\Delta C_{pk,j}$ ranges from $-0.645$ to $+0.298$, with mean absolute distortion $\overline{|\Delta C_{pk}|}\approx 0.18$. Among the 10 non-normal dimensions, 9 exhibit $|\Delta C_{pk}|>0.15$, indicating that model choice alone can introduce substantial variation in capability estimates, which becomes amplified in downstream defect-risk evaluation.

The direction of distortion varies across dimensions: large negative values indicate that the normal assumption overstates capability under skewed or heavy-tailed behavior, whereas positive values indicate conservative bias. More importantly, these distortions are asymmetric and distribution-dependent, reflecting structural sensitivity of capability indices to tail behavior.

\begin{figure}[htbp]
	\centerline{\includegraphics[width=1.00\linewidth]{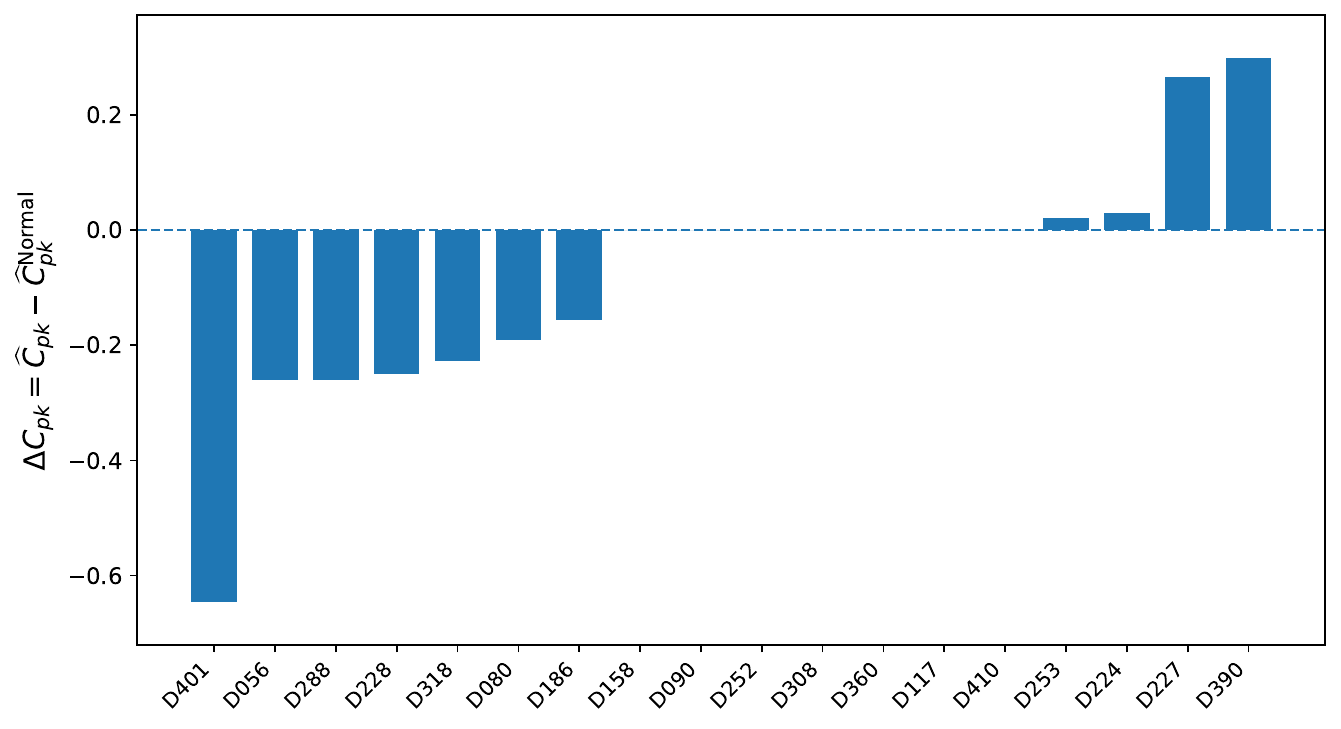}}
	\caption{Model-induced capability distortion 
		$\Delta C_{pk} = \widehat{C}_{pk} - \widehat{C}_{pk}^{\mathrm{Normal}}$
		across the 18 selected dimensions. Negative values indicate inflation of capability under the normal assumption relative to the fitted distribution; positive values indicate conservative bias. The observed distortions, while moderate in capability space, can lead to amplified differences in defect-risk metrics and decision outcomes.}
	\label{fig:val_fig2_delta_model}
\end{figure}

Although Eq.~\eqref{eq:delta_model_val} captures differences in capability estimates, their impact on decision-making is governed by nonlinear amplification. Small model-induced shifts in $C_{pk}$ can translate into disproportionately large changes in defect-risk metrics, particularly near the capability threshold. Consequently, model choice introduces an additional source of variability that is amplified in defect-risk space and directly affects decision outcomes.

These results show that amplification is driven not only by finite-sample variability but also by model uncertainty. Even modest model-induced differences in capability estimates can be significantly magnified in defect-risk space, leading to materially different decision outcomes. This establishes model choice as a structurally amplified source of decision instability in capability-based analysis.

%%%%%%%%%%%%%%%%%%%%%%%%%%%%%%%%%%%%%%%%%%%%%%%%%%
\subsection{Nonlinear Amplification under Model Uncertainty}
\label{sec:val_sec5}

We empirically validate the nonlinear amplification mechanism by examining how finite-sample dispersion variability propagates into defect-risk metrics. For each dimension, defect probability and PPM are evaluated under both normal and fitted distributions, as summarized in Table~\ref{tab:val_stats18}.

At $n=32$, Eq.~\eqref{eq:cvS_approx} gives $\CV(S)\approx 0.127$, indicating non-negligible dispersion variability even under ideal normal sampling. To assess its impact, we fix $\bar{X}$ and evaluate defect rates at perturbed dispersion levels
\[
S_{+}=S(1+\CV), \qquad
S_{-}=S(1-\CV).
\]
The resulting values $\widehat{\PPM}_0$, $\widehat{\PPM}_{+}$, and $\widehat{\PPM}_{-}$ are reported in Table~\ref{tab:val_stats18}.

The empirical patterns reveal pronounced nonlinear amplification across regimes. In the mid-capability regime (e.g., D056, D080, D158, D186, D224, D288), a $\pm 12.7\%$ perturbation in dispersion produces relative changes in predicted PPM that substantially exceed $12.7\%$. In the low-capability regime (e.g., D360 and D390), baseline defect rates are already large, and perturbations in $S$ generate significant absolute changes in PPM. In the high-capability regime (e.g., D308, D318, D401), baseline $\widehat{\PPM}_0$ is near zero, yet upward perturbations in $S$ produce orders-of-magnitude increases in defect risk on a relative scale.

A systematic discrepancy is also observed between fitted-distribution-based and normal-based defect rates. For several dimensions (e.g., D056, D186, D401), fitted PPM values exceed their normal-based counterparts by multiple orders of magnitude, indicating that the normal assumption can substantially underestimate defect risk in the presence of skewness or heavy tails.

These results provide direct empirical validation of the nonlinear amplification mechanism: modest variability in dispersion, though appearing limited in capability space, can induce disproportionately large changes in defect-risk metrics, an effect that is further intensified under model uncertainty.

%%%%%%%%%%%%%%%%%%%%%%%%%%%%%%%%%%%%%%%%%%%%%%%%%%
\subsection{Decision Instability and Boundary-Focused Evidence}
\label{sec:val_boundary}

The final question is whether amplification-induced instability has practical consequences for real decision making. Table~\ref{tab:val_stats18} shows that bootstrap pass probabilities $\widehat{p}$ and lower confidence bounds $\LCB_{0.95}$ vary substantially across the 18 selected dimensions, despite a common sample size $n=32$. High-capability cases (e.g., D308, D318) remain stably acceptable, while low-capability cases (e.g., D360, D390) remain stably rejected. In contrast, dimensions near the decision threshold exhibit intermediate pass probabilities and lower confidence bounds below $C_0$, indicating that deterministic point estimates do not reliably capture decision outcomes.

This behavior is consistent with the amplification framework: near the threshold, modest estimation variability can be magnified into materially different decision outcomes. Reliability-aware summaries such as $\widehat{p}$ and $\LCB_{0.95}$ therefore provide a practical correction to deterministic thresholding by distinguishing stable decisions from amplification-sensitive ones.

To assess how frequently such boundary-sensitive regimes arise, we analyze the full 500-dimension dataset using reported $\widehat{C}_{pk}^{\mathrm{normal}}$ values. The empirical concentration around the release threshold $C_0=1.33$ is substantial:
\[
\begin{aligned}
	&15 \text{ dimensions satisfy } |\widehat{C}_{pk}^{\mathrm{normal}}-1.33|\le 0.01,\\
	&32 \text{ satisfy } |\widehat{C}_{pk}^{\mathrm{normal}}-1.33|\le 0.02,\\
	&80 \text{ satisfy } |\widehat{C}_{pk}^{\mathrm{normal}}-1.33|\le 0.05,\\
	&151 \text{ satisfy } |\widehat{C}_{pk}^{\mathrm{normal}}-1.33|\le 0.10.
\end{aligned}
\]

Defining $d_j = |\widehat{C}_{pk,j}^{\mathrm{normal}}-C_0|$ and examining the 20 closest cases yields a range of $1.309 \le \widehat{C}_{pk,j}^{\mathrm{normal}} \le 1.339$, with mean $1.3269$, placing all within $\pm 0.021$ of the threshold. 

These results demonstrate that a substantial fraction of real manufacturing dimensions operate in a narrow neighborhood of the decision boundary. In such regimes, classification is intrinsically fragile: small perturbations in estimated capability can alter pass/fail outcomes. Amplification-induced decision instability is therefore not a rare edge case, but a practically pervasive feature of capability-based decision systems.

Collectively, the empirical evidence supports three conclusions: (i) model choice can materially distort reported capability, (ii) finite-sample variability is nonlinearly amplified in defect-risk space, and (iii) because many real dimensions lie near capability thresholds, this amplification leads to intrinsically fragile classification and has direct, unavoidable consequences for decision reliability in real manufacturing systems.

%%%%%%%%%%%%%%%%%%%%%%%%%%%%%%%%%%%%%%%%%%%%%%%%%%%%%%%%%%%%%%%%%%%%
\section{Conclusion}
\label{sec:conclusion}

Process capability indices such as $C_{pk}$ are widely used in manufacturing quality management to assess whether a process satisfies engineering specification requirements. In practice, capability approval is typically implemented through deterministic threshold rules applied to estimated indices. Because these estimators are computed from finite samples, the resulting approval decisions are inherently stochastic. This paper shows that such instability cannot be explained by estimator variability alone, but is driven by a nonlinear amplification mechanism—quantified through a closed-form coefficient—that transforms capability uncertainty into defect-risk space. While capability indices depend approximately linearly on dispersion estimation, defect probability is governed by tail curvature. Consequently, small estimation errors can be disproportionately amplified when translated into defect-risk measures such as parts-per-million (PPM), leading to substantial decision instability, particularly near threshold regions.

Building on this insight, capability approval is reformulated as a reliability-aware decision problem, in which approval requires that the probability of satisfying the capability requirement exceeds a specified confidence level. Within this framework, sample size plays a central role not only in reducing estimator variability, but in attenuating amplification-induced instability. Reliability-based criteria therefore provide a principled approach to controlling misclassification risk and achieving stable decision outcomes.

Empirical validation using large-scale manufacturing data confirms that model specification, finite-sample dispersion variability, and nonlinear tail sensitivity jointly influence capability-based decisions under realistic conditions. In particular, the prevalence of near-threshold dimensions demonstrates that amplification-induced instability is not exceptional but pervasive in practice. Overall, this work establishes a unified framework linking capability estimation, nonlinear amplification, reliability-aware decision rules, and sample-size planning, with amplification explicitly quantified through a closed-form sensitivity measure. By explicitly accounting for the amplification of uncertainty, the proposed approach provides a more robust foundation for capability-based decision-making in modern manufacturing environments. Furthermore, the results highlight that distributional misspecification can significantly distort defect-risk estimates, reinforcing the need for model-aware capability evaluation.

%%%%%%%%%%%%%%%%%%%%%%%%%%%%%%%%%%%%%%%%%%%%%%%%%%%%%%%%%%%%%%%%%%%%%%
\appendix
\numberwithin{equation}{section}
\numberwithin{figure}{section}
\section*{APPENDIX}
\section{Raw Data}
\label{app:raw-data}

The raw measurement data corresponding to the 18 selected dimensions analyzed in Section~\ref{sec:empirical} are provided in tabular form. These data include the original observations used for capability estimation, bootstrap analysis, and empirical validation of the nonlinear amplification effects.
\begin{table*}[!t]
	\centering
	\footnotesize
	\renewcommand{\arraystretch}{1.0}
	\setlength{\tabcolsep}{1.65pt}
	\resizebox{\textwidth}{!}{%
		\begin{tabular}{ccccccccccccccccccc}
			\toprule
			NO. & D056 & D080 & D090 & D117 & D158 & D186 & D224 & D227 & D228 & D252 & D253 & D288 & D308 & D318 & D360 & D390 & D401 & D410 \\
			\midrule
			$T$ & 0.55 & 1.17 & 1.65 & 2.27 & 2.74 & 3.28 & 3.70 & 4.21 & 5.10 & 5.52 & 6.16 & 6.65 & 7.13 & 7.78 & 8.20 & 8.65 & 9.19 & 9.87 \\
			$Tol+$ & 0.10 & 0.10 & 0.05 & 0.10 & 0.10 & 0.10 & 0.05 & 0.10 & 0.05 & 0.25 & 0.05 & 0.05 & 0.10 & 0.05 & 0.10 & 0.10 & 0.05 & 0.08 \\
			$Tol-$ & 0.10 & 0.10 & 0.05 & 0.10 & 0.10 & 0.10 & 0.05 & 0.10 & 0.05 & 0.25 & 0.05 & 0.05 & 0.10 & 0.05 & 0.10 & 0.10 & 0.05 & 0.08 \\
			\midrule
			1 & 0.534 & 1.209 & 1.662 & 2.299 & 2.691 & 3.251 & 3.682 & 4.230 & 5.073 & 5.567 & 6.124 & 6.617 & 7.139 & 7.782 & 8.232 & 8.721 & 9.197 & 9.834 \\
			2 & 0.525 & 1.184 & 1.664 & 2.263 & 2.706 & 3.318 & 3.682 & 4.214 & 5.079 & 5.636 & 6.120 & 6.623 & 7.148 & 7.780 & 8.298 & 8.738 & 9.182 & 9.857 \\
			3 & 0.476 & 1.174 & 1.628 & 2.263 & 2.738 & 3.245 & 3.715 & 4.234 & 5.083 & 5.553 & 6.129 & 6.632 & 7.153 & 7.766 & 8.145 & 8.689 & 9.197 & 9.850 \\
			4 & 0.517 & 1.155 & 1.641 & 2.316 & 2.707 & 3.294 & 3.690 & 4.238 & 5.088 & 5.648 & 6.120 & 6.620 & 7.164 & 7.769 & 8.230 & 8.738 & 9.197 & 9.832 \\
			5 & 0.515 & 1.188 & 1.646 & 2.302 & 2.693 & 3.292 & 3.695 & 4.286 & 5.085 & 5.603 & 6.124 & 6.626 & 7.149 & 7.784 & 8.114 & 8.724 & 9.197 & 9.849 \\
			6 & 0.518 & 1.164 & 1.622 & 2.258 & 2.732 & 3.298 & 3.679 & 4.236 & 5.088 & 5.468 & 6.122 & 6.622 & 7.155 & 7.775 & 8.194 & 8.676 & 9.176 & 9.853 \\
			7 & 0.541 & 1.226 & 1.649 & 2.284 & 2.763 & 3.310 & 3.687 & 4.238 & 5.085 & 5.577 & 6.126 & 6.625 & 7.144 & 7.793 & 8.216 & 8.691 & 9.186 & 9.821 \\
			8 & 0.511 & 1.171 & 1.649 & 2.257 & 2.752 & 3.310 & 3.694 & 4.234 & 5.080 & 5.621 & 6.127 & 6.626 & 7.170 & 7.779 & 8.283 & 8.661 & 9.201 & 9.867 \\
			9 & 0.527 & 1.161 & 1.634 & 2.257 & 2.745 & 3.311 & 3.694 & 4.237 & 5.099 & 5.567 & 6.119 & 6.619 & 7.155 & 7.783 & 8.247 & 8.722 & 9.194 & 9.862 \\
			10 & 0.512 & 1.189 & 1.642 & 2.319 & 2.705 & 3.293 & 3.694 & 4.233 & 5.105 & 5.623 & 6.120 & 6.631 & 7.148 & 7.781 & 8.293 & 8.729 & 9.176 & 9.836 \\
			11 & 0.487 & 1.169 & 1.629 & 2.258 & 2.697 & 3.260 & 3.702 & 4.231 & 5.079 & 5.614 & 6.128 & 6.622 & 7.160 & 7.782 & 8.292 & 8.743 & 9.193 & 9.831 \\
			12 & 0.484 & 1.158 & 1.645 & 2.298 & 2.738 & 3.310 & 3.693 & 4.234 & 5.087 & 5.528 & 6.125 & 6.634 & 7.150 & 7.777 & 8.266 & 8.735 & 9.194 & 9.852 \\
			13 & 0.512 & 1.201 & 1.634 & 2.262 & 2.737 & 3.299 & 3.695 & 4.235 & 5.112 & 5.569 & 6.120 & 6.621 & 7.155 & 7.777 & 8.212 & 8.724 & 9.197 & 9.848 \\
			14 & 0.523 & 1.165 & 1.629 & 2.315 & 2.721 & 3.277 & 3.717 & 4.303 & 5.093 & 5.632 & 6.128 & 6.622 & 7.159 & 7.773 & 8.182 & 8.722 & 9.194 & 9.867 \\
			15 & 0.475 & 1.149 & 1.645 & 2.264 & 2.719 & 3.278 & 3.694 & 4.236 & 5.084 & 5.521 & 6.119 & 6.616 & 7.154 & 7.771 & 8.197 & 8.741 & 9.196 & 9.862 \\
			16 & 0.531 & 1.202 & 1.666 & 2.303 & 2.697 & 3.282 & 3.705 & 4.232 & 5.085 & 5.603 & 6.121 & 6.619 & 7.160 & 7.773 & 8.265 & 8.731 & 9.189 & 9.819 \\
			17 & 0.526 & 1.151 & 1.648 & 2.262 & 2.716 & 3.299 & 3.715 & 4.234 & 5.099 & 5.579 & 6.120 & 6.628 & 7.154 & 7.774 & 8.214 & 8.703 & 9.181 & 9.835 \\
			18 & 0.535 & 1.215 & 1.651 & 2.313 & 2.720 & 3.237 & 3.713 & 4.235 & 5.099 & 5.614 & 6.128 & 6.621 & 7.151 & 7.774 & 8.293 & 8.724 & 9.176 & 9.828 \\
			19 & 0.522 & 1.166 & 1.655 & 2.276 & 2.721 & 3.309 & 3.697 & 4.239 & 5.080 & 5.574 & 6.119 & 6.631 & 7.150 & 7.777 & 8.298 & 8.716 & 9.193 & 9.826 \\
			20 & 0.528 & 1.189 & 1.646 & 2.291 & 2.750 & 3.285 & 3.681 & 4.198 & 5.086 & 5.601 & 6.124 & 6.636 & 7.151 & 7.788 & 8.250 & 8.739 & 9.203 & 9.862 \\
			21 & 0.530 & 1.164 & 1.669 & 2.288 & 2.724 & 3.299 & 3.705 & 4.234 & 5.097 & 5.550 & 6.129 & 6.623 & 7.151 & 7.777 & 8.266 & 8.742 & 9.181 & 9.827 \\
			22 & 0.527 & 1.243 & 1.653 & 2.221 & 2.737 & 3.296 & 3.680 & 4.234 & 5.085 & 5.559 & 6.127 & 6.621 & 7.167 & 7.773 & 8.279 & 8.734 & 9.194 & 9.845 \\
			23 & 0.531 & 1.160 & 1.632 & 2.266 & 2.738 & 3.310 & 3.685 & 4.237 & 5.083 & 5.660 & 6.124 & 6.618 & 7.171 & 7.775 & 8.195 & 8.745 & 9.193 & 9.847 \\
			24 & 0.531 & 1.182 & 1.639 & 2.255 & 2.699 & 3.292 & 3.718 & 4.239 & 5.088 & 5.576 & 6.122 & 6.620 & 7.143 & 7.779 & 8.227 & 8.727 & 9.195 & 9.844 \\
			25 & 0.521 & 1.184 & 1.648 & 2.275 & 2.726 & 3.311 & 3.693 & 4.239 & 5.071 & 5.573 & 6.121 & 6.628 & 7.161 & 7.773 & 8.196 & 8.731 & 9.204 & 9.841 \\
			26 & 0.532 & 1.184 & 1.654 & 2.266 & 2.756 & 3.322 & 3.688 & 4.232 & 5.080 & 5.568 & 6.124 & 6.623 & 7.153 & 7.772 & 8.258 & 8.747 & 9.169 & 9.844 \\
			27 & 0.536 & 1.169 & 1.655 & 2.293 & 2.744 & 3.300 & 3.709 & 4.234 & 5.080 & 5.638 & 6.120 & 6.642 & 7.150 & 7.776 & 8.293 & 8.699 & 9.200 & 9.853 \\
			28 & 0.529 & 1.181 & 1.661 & 2.255 & 2.718 & 3.303 & 3.689 & 4.234 & 5.085 & 5.522 & 6.119 & 6.629 & 7.154 & 7.777 & 8.221 & 8.724 & 9.195 & 9.836 \\
			29 & 0.517 & 1.161 & 1.646 & 2.290 & 2.690 & 3.307 & 3.715 & 4.223 & 5.090 & 5.524 & 6.122 & 6.621 & 7.158 & 7.771 & 8.285 & 8.734 & 9.193 & 9.844 \\
			30 & 0.537 & 1.149 & 1.651 & 2.255 & 2.692 & 3.238 & 3.708 & 4.234 & 5.087 & 5.516 & 6.120 & 6.624 & 7.147 & 7.803 & 8.166 & 8.714 & 9.184 & 9.857 \\
			31 & 0.506 & 1.171 & 1.642 & 2.278 & 2.742 & 3.279 & 3.711 & 4.234 & 5.099 & 5.492 & 6.120 & 6.621 & 7.151 & 7.782 & 8.195 & 8.686 & 9.191 & 9.853 \\
			32 & 0.509 & 1.151 & 1.648 & 2.286 & 2.749 & 3.285 & 3.693 & 4.274 & 5.082 & 5.605 & 6.126 & 6.620 & 7.160 & 7.777 & 8.282 & 8.689 & 9.182 & 9.846 \\
			\bottomrule
		\end{tabular}%
	}
	\caption{Raw data ($n=32$) for the 18 selected dimensions, with nominal values and upper/lower tolerances included.}
	\label{tab:val_rawdata18}
\end{table*}

%%%%%%%%%%%%%%%%%%%%%%%%%%%%%%%%%%%%%%%%%%%%%%%
\section*{DECLARATION}
\begin{description}
	\item[Funding:] This study received no external funding.
	
	\item[Conflicts of interest:] The authors declare no conflicts of interest.
	
	\item[Availability of data and material:] The empirical dataset is derived from anonymized modified manufacturing data. Processed data are available from the corresponding author upon reasonable request.
	
	\item[Code availability:] Simulation and analysis code are available from the corresponding author upon reasonable request.
	
	\item[Ethics approval:] Not applicable.
	
	\item[Consent for publication:] All authors approve the final manuscript.
\end{description}
%%%%%%%%%%%%%%%%%%%%%%%%%%%%%%%%%%%%%%%%%%%%%%%
 % Here's where you specify the bibliography style file.
 % The full file name for the bibliography style file 
 \bibliographystyle{unsrtnat}
 \bibliography{references}
 %%%%%%%%%%%%%%%%%%%%%%%%%%%%%%%%%%%%%%%%%%%%%%%
 \end{document}